\newcounter{myctr}
\def\myitem{\refstepcounter{myctr}\bibfont\noindent\ifnum\themyctr>9\else\phantom{0}\fi\hangindent17pt\themyctr.\enskip}

\documentclass{article}
\usepackage[english]{babel}
\usepackage{amsmath,amssymb,graphicx,hyperref,makeidx}
\makeindex

\newcommand{\Mu}{\mathrm{M}}
\newcommand{\mathd}{\mathrm{d}}
\newcommand{\mathe}{\mathrm{e}}

\newcommand{\tmem}[1]{{\em #1\/}}
\newcommand{\tmop}[1]{\ensuremath{\operatorname{#1}}}
\newcommand{\tmrsub}[1]{\ensuremath{_{\textrm{#1}}}}
\newcommand{\tmtextbf}[1]{\text{{\bfseries{#1}}}}

\begin{document}
\title{Comparison of distances and\\
  entropic distinguishability quantifiers\\
  for the detection of memory effects}

\author{Bassano Vacchini\\
Dipartimento di Fisica ``Aldo Pontremoli'', \\
Universit{\`a}
  degli Studi di Milano\\
  Via Celoria 16, 20133 Milan, Italy\\
  Istituto Nazionale di Fisica Nucleare, Sezione di Milano\\
  Via Celoria 16, 20133 Milan, Italy\\
  bassano.vacchini@mi.infn.it}

\maketitle

\begin{abstract}
  We consider a recently introduced framework for the description of memory
  effects based on quantum state distinguishability quantifiers, in which
  entropic quantifiers can be included. After briefly presenting the approach,
  we validate it considering the performance of different quantifiers in the
  characterization of the reduced dynamics of a two-level system undergoing
  decoherence. We investigate the different behavior of these quantifiers in
  the dependence on physical features of the model, such as environmental
  temperature and coupling strength. It appears that the performance of the
  different quantifiers conveys the same physical information, though with
  different sensitivities, thus supporting robustness of the approach.
\end{abstract}

\section{Introduction}

The characterization of open quantum system dynamics in view of memory effects
has recently attracted a lot of attention, and different approaches have been
pursued in this direction, addressing the very question of what is a
non-Markovian quantum process
{\cite{Rivas2014a,Breuer2016a,Devega2017a,Li2018a,Milz2021a,Chruscinski2022a}}.
We here focus on a strategy introduced in the seminal paper
{\cite{Breuer2009b}}, that only requires knowledge of the reduced state of the
open system as a function of time. The approach was initially formulated
relying on the trace distance to compare the evolution of different initial
system states. It was later shown that also entropic quantifiers based on the
quantum relative entropy can be considered {\cite{Megier2021a,Smirne2022b}}.
In the present manuscript we want to investigate the different behavior of
these quantifiers, to check whether indeed the thus obtained notion of
non-Markovian dynamics is robust with respect to the considered quantifier,
provided it satisfies some natural general properties. To this aim we
investigate the measure of non-Markovianity introduced in
{\cite{Breuer2009b}}, for the case of a spin-boson dephasing model, evaluating
it for different quantifiers. In particular we study its dependence on
physical parameters of the model, such as temperature of the bosonic bath or
coupling strength. All quantifiers exhibit a physically coherent behavior,
though showing different sensitivities. In particular, distances have a more
marked dependence on the physical parameters of the model with respect to the
considered divergences.

\section{A framework for the characterization of \ quantum reduced dynamics}

We first introduce the framework that we plan to use to provide a
characterization of a reduced quantum dynamics in view of memory effects. It
is essentially based on the approach originated from {\cite{Breuer2009b}},
though formulated so as to include entropic distinguishability quantifiers,
that do not obey the standard triangle inequality. The emphasis is on the
association of memory with information that is initially stored outside the
open system, in well-identified degrees of freedom, and later retrievable by
performing measurements on the system only.

\subsection{States comparison and Markov condition}\label{sect:markov}

Following {\cite{Smirne2022b}} we denote with $\mathfrak{S} (\rho, \sigma)$ a
quantifier of the distinguishability between two statistical operators $\rho$
and $\sigma$. In the first instance we assume symmetry
\begin{eqnarray}
  \mathfrak{S} (\rho, \sigma) & = & \mathfrak{S} (\sigma, \rho), 
  \label{eq:symm}
\end{eqnarray}
and boundedness
\begin{equation}
  0 \leqslant \mathfrak{S} (\rho, \sigma) \leqslant 1, \label{eq:bound}
\end{equation}
together with perfect discrimination capability for the case of orthogonal
states
\begin{equation}
  \mathfrak{S} (\rho, \sigma) = 1 \quad \Leftrightarrow \quad \rho \perp 
  \sigma \label{eq:normorto}
\end{equation}
and
\begin{equation}
  \mathfrak{S} (\rho, \sigma) = 0 \quad \Leftrightarrow \quad \rho = \sigma
  \label{eq:indid} .
\end{equation}
We further ask the key property of being a contraction with respect to the
action of a completely positive trace preserving map $\Phi$, describing a
well-defined evolution
\begin{eqnarray}
  \mathfrak{S} (\Phi [\rho], \Phi [\sigma]) & \leqslant & \mathfrak{S} (\rho,
  \sigma) .  \label{eq:cpcontra}
\end{eqnarray}
Finally, for the sake of the desired association between memory effects and
information exchange, we need a generalization of the triangle inequality,
formulated as follows
\begin{eqnarray}
  \mathfrak{S} (\sigma, \rho) -\mathfrak{S} (\tau, \rho) & \leqslant & \phi
  (\mathfrak{S}(\sigma, \tau))  \label{eq:tlike}
\end{eqnarray}
with $\tau$ an arbitrary state and $\phi$ a positive concave function starting
from the value zero at the origin. The requirements of symmetry and
boundedness, let alone the triangle inequality, are clearly not obviously
satisfied by entropic quantifiers. As we shall see, however, the enforcement
of boundedness allows to symmetrize and to derive a suitable triangle-like
inequality.

Given a distinguishability quantifier satisfying the above properties
(\ref{eq:symm}) to (\ref{eq:tlike}), we define a reduced dynamics described by
a collection of completely positive trace preserving maps $\Phi (t)$ according
to
\begin{eqnarray}
  \rho_S (t) & = & \Phi (t) \rho_S (0) 
\end{eqnarray}
to be Markovian if
\begin{eqnarray}
  \mathfrak{S} (\rho_S^1 (t), \rho_S^2 (t)) & \leqslant & \mathfrak{S}
  (\rho_S^1 (s), \rho_S^2 (s)) \qquad \forall t \geqslant s \geqslant 0, 
  \label{eq:monotoneq}
\end{eqnarray}
for any pair of initial conditions $\rho_S^1 (0)$ and $\rho_S^2 (0)$.
According to this definition a quantum reduced dynamics is said to be
Markovian if the distinguishability between two states is a monotonically
non-increasing function of time. It is easily checked that this condition is
verified for the case of quantum dynamical semigroups, so that the standard
identification of quantum Markovian processes complies with this definition,
that crucially depends on validity of Eq.~(\ref{eq:cpcontra}). The violation
of this condition identifies reduced dynamics which are non-Markovian.

\subsection{Distances and entropic quantifiers}\label{sect:quantifiers}

We will consider essentially two situations. One the one hand, the case in
which the quantifier $\mathfrak{S}$ is a distance in the mathematical sense,
so that only contractivity under completely positive trace preserving maps has
to be checked. In particular, this is true for the trace distance based on the
natural norm on the space of trace class operators, that is the trace norm
\begin{eqnarray}
  \| A \|_1 & = & \tmop{Tr} | A | .  \label{eq:norma1}
\end{eqnarray}
On the other hand, the case in which a quantifier $\mathfrak{S}$ with the
desired properties is obtained starting from the quantum relative entropy,
defined for positive operators as
\begin{eqnarray}
  S (A, B) & = & \tmop{Tr} A (\ln A - \ln B) .  \label{eq:qre}
\end{eqnarray}
In this case, contractivity under completely positive trace preserving maps is
ensured, but further elaboration is needed in order to warrant boundedness and
a variant of the triangle inequality {\cite{Smirne2022b}}. The main aim of
this paper is to benchmark the behavior of these two families with respect to
the insurgence of non-Markovian dynamics and its dependence on physical
features of the model.

\subsubsection{Trace distance}

We first recall the definition of trace distance between statistical operators
\begin{eqnarray}
  D (\rho, \sigma) & = & \frac{1}{2} \| \rho - \sigma \|_1,  \label{eq:dtrace}
\end{eqnarray}
where the factor $\tfrac{1}{2}$ comes from the requirement
Eq.~(\ref{eq:bound}), contractivity can be directly proven
{\cite{Ruskai1994a}} and the other properties follow from the fact that
Eq.~(\ref{eq:norma1}) is a norm. In particular, the triangle inequality
warrants Eq.~(\ref{eq:tlike}) with $\phi$ the identity function $\phi (x) =
x$. This distinguishability quantifier naturally arises in a discrimination
task and was the first introduced in the study of non-Markovian dynamics
{\cite{Breuer2009b}}.

\subsubsection{Entropic quantifiers}

We now introduce a variant of the quantum relative entropy, that satisfies all
the above mentioned properties. We consider the quantity
\begin{equation}
  S  (\rho, \mu \rho + (1 - \mu) \sigma), \label{eq:relskew}
\end{equation}
with $0 < \mu < 1$, first introduced in
{\cite{Audenaert2011xxx,Audenaert2014a}} and initially called telescopic
relative entropy. It can be shown that this quantity takes values in the range
\begin{equation}
  0 \leqslant S  (\rho, \mu \rho + (1 - \mu) \sigma) \leqslant \ln (1 / \mu),
\end{equation}
and most importantly it obeys the triangle-like inequalities
\begin{equation}
  S (\sigma, \mu \sigma + (1 - \mu) \rho_1) - S (\sigma, \mu \sigma + (1 -
  \mu) \rho_2) \leqslant \ln \left( 1 + \frac{1 - \mu}{\mu} D (\rho_1, \rho_2)
  \right) \label{eq:b1}
\end{equation}
and
\begin{equation}
  S (\rho_1, \mu \rho_1 + (1 - \mu) \sigma) - S (\rho_2, \mu \rho_2 + (1 -
  \mu) \sigma) \leqslant D (\rho_1, \rho_2) \ln \left( 1 + \frac{1 - \mu}{\mu}
  \frac{1}{D (\rho_1, \rho_2)} \right) \label{eq:b2}
\end{equation}
where $D$ denotes the trace distance introduced in Eq.~(\ref{eq:dtrace}).
These properties allow to consider symmetric versions of the quantity and to
fix the desired range Eq.~(\ref{eq:bound}). There are two natural choices that
can be considered {\cite{Smirne2022b}}. A first immediate choice is given by
the quantum skew divergence
\begin{eqnarray}
  S_{\mu} (\rho, \sigma) & = & \frac{\mu}{\ln (1 / \mu)} S (\rho, \mu \rho +
  (1 - \mu) \rho) \nonumber\\
  &  & + \frac{(1 - \mu)}{\ln (1 / (1 - \mu))} S (\sigma, (1 - \mu) \sigma +
  \mu \rho),  \label{eq:qskew}
\end{eqnarray}
symmetric by construction under the exchange
\begin{equation}
  \mu \leftrightarrow 1 - \mu \qquad \rho \leftrightarrow \sigma .
  \label{eq:simmetria}
\end{equation}
Thanks to Eq.~(\ref{eq:b1}) and Eq.~(\ref{eq:b2}) we further have
\begin{eqnarray}
  S_{\mu} (\rho, \sigma) - S_{\mu} (\rho, \tau) & \leqslant & \varsigma_{\mu}
  \sqrt[4]{S_{\mu} (\sigma, \tau)},  \label{eq:tlikeS}
\end{eqnarray}
so that Eq.~(\ref{eq:tlike}) is satisfied with $\phi (x) = \varsigma_{\mu}
\sqrt[4]{x}$ and
\begin{eqnarray}
  \varsigma_{\mu} & = & \ln \left( \frac{1}{\mu (1 - \mu)} \right)
  \sqrt[4]{\frac{\mu (1 - \mu)}{2 H (\{ \mu, 1 - \mu \}) \ln^3 (\mu) \ln^3 (1
  - \mu)}} . 
\end{eqnarray}
In Eq.~(\ref{eq:qskew}) each term is separately normalized. Another more
subtle choice is given by the expression
\begin{eqnarray}
  K_{\mu} (\rho, \sigma) & = & \frac{\mu}{H (\{ \mu, 1 - \mu \})} S (\rho, \mu
  \rho + (1 - \mu) \sigma) \nonumber\\
  &  & + \frac{1 - \mu}{H (\{ 1 - \mu, \mu \})} S (\sigma, (1 - \mu) \sigma +
  \mu \rho),  \label{eq:hskew}
\end{eqnarray}
that still shares the symmetry Eq.~(\ref{eq:simmetria}) and corresponds to a
normalized version of the Holevo information or Holevo $\chi$ quantity for the
case of an ensemble composed of two states only. We have the simple identity
\begin{eqnarray}
  K_{\mu} (\rho, \sigma) & = & \frac{\chi  (\{ \mu, \rho ; 1 - \mu, \sigma
  \})}{H (\{ 1 - \mu, \mu \})}, 
\end{eqnarray}
so that we will call the distinguishability quantifier $K_{\mu} (\rho,
\sigma)$ Holevo skew divergence. This quantity is still bounded and in the
range Eq.~(\ref{eq:bound}), furthermore it obeys the inequality
\begin{eqnarray}
  K_{\mu} (\rho, \sigma) - K_{\mu} (\rho, \tau) & \leqslant & \kappa_{\mu}
  \sqrt[4]{K_{\mu} (\sigma, \tau)},  \label{eq:tlikeK}
\end{eqnarray}
with
\begin{eqnarray}
  \kappa_{\mu} & = & \sqrt[4]{\frac{8 \mu (1 - \mu)}{H (\{ \mu, 1 - \mu
  \})^3}} . 
\end{eqnarray}
We will see that they have a very close performance in the characterization of
non-Markovian dynamics. In both cases, the crucial contractivity property
Eq.~(\ref{eq:cpcontra}) is warranted by contractivity of the quantum relative
entropy under the action of completely positive trace preserving maps
{\cite{Lindblad1975a}}.

\subsubsection{Jensen-Shannon divergence}

A special role is played by the case $\mu = \tfrac{1}{2}$, indeed we have
\begin{equation}
  S_{1 / 2} (\rho, \sigma) = K_{1 / 2} (\rho, \sigma) = J (\rho, \sigma)
  \label{eq:skj}
\end{equation}
where $J (\rho, \sigma)$ denotes the so-called Jensen-Shannon divergence
\begin{eqnarray}
  J (\rho, \sigma) & = & \frac{1}{2 \ln 2} \left[ S \left( \rho, \frac{\rho +
  \sigma}{2} \right) + S \left( \sigma, \frac{\rho + \sigma}{2} \right)
  \right] .  \label{eq:j}
\end{eqnarray}
It is worth mentioning that its square root provides a distance in the strict
mathematical sense {\cite{Sra2021a,Virosztek2021a}}, so that
\begin{eqnarray}
  \sqrt{J (\rho, \sigma)} - \sqrt{J (\rho, \tau)} & \leqslant & \sqrt{J
  (\sigma, \tau)},  \label{eq:tlikebis}
\end{eqnarray}
while symmetry is already explicit in the very expression Eq.~(\ref{eq:j}).

We are thus led to consider two distances, namely trace distance $D$ and
square root of the Jensen-Shannon divergence $\sqrt{J}$, as well as two
entropic quantifiers, Holevo and quantum skew divergence, which we will both
take for the value $\mu = 1 / 4$.

\subsection{Measure of distinguishability revivals}\label{sect:markov-measure}

The definition Eq.~(\ref{eq:monotoneq}) naturally suggests a way to quantify
the violation of the Markovian condition, as considered in
{\cite{Breuer2009b}} introducing a so-called non-Markovianity measure. We thus
introduce the adimensional quantity
\begin{eqnarray}
  \Mu (\Phi (t), \mathfrak{S}, \rho^{1, 2}_S (0)) & = &
  \int_{\dot{\mathfrak{S}} > 0} \mathd t \dot{\mathfrak{S}} (\rho_S (t),
  \sigma_S (t)),  \label{eq:nm-pair}
\end{eqnarray}
where the integration is restricted to the regions in which the integrand
$\dot{\mathfrak{S}} (\rho_S (t), \sigma_S (t))$ is positive, so that it can be
equivalently written as
\begin{eqnarray}
  \Mu (\Phi (t), \mathfrak{S}, \rho^{1, 2}_S (0)) & = & \sum_n
  [\mathfrak{S}(\rho^1_S (t^n_f), \rho^2_S (t^n_f)) -\mathfrak{S}(\rho^1_S
  (t^n_i), \rho^2_S (t^n_i))],  \label{eq:nm-sum}
\end{eqnarray}
with $t^n_i$ and $t^n_f$ initial and final time of the $n$-th revival.
Following {\cite{Breuer2009b,Laine2010a}} we therefore define the
quantity{\index{$\Mu (\Phi, \mathfrak{S})$}}
\begin{eqnarray}
  \Mu (\Phi (t), \mathfrak{S}) & = & \sup_{\rho^{1, 2}_S (0)} \Mu (\Phi (t),
  \mathfrak{S}, \rho^{1, 2}_S (t_0))  \label{eq:nm-measure}
\end{eqnarray}
as measure of non-Markovianity associated to the evolution described by the
time dependent collection of completely positive trace preserving maps $\Phi
(t)$. The basic idea is to consider the sum of the revivals in
distinguishability, depending on the considered quantifier $\mathfrak{S}$.
Given that the existence and the amount of the revivals depends on the pair of
initial states whose evolution has to be compared, the measure is defined
according to Eq.~(\ref{eq:nm-measure}) by optimizing over the initial pair.
Indeed, the same environment differently affects distinct initial system
states, and it is therefore important to explore the initial state dependence.
While the value itself of $\Mu (\Phi (t), \mathfrak{S})$ does not have a
special meaning, the dependence of this non-Markovianity measure on physical
parameters of the model does provide interesting information on the origin of
memory effects.

\subsection{Interpretation in terms of information
backflow}\label{sect:info-flow}

We now want to provide a motivation for the definition of quantum Markovian
dynamics given in Sect.~\ref{sect:markov}, building on validity of the
condition Eq.~(\ref{eq:tlike}). The very introduction of a reduced dynamics
was possible with reference to a bipartition of the Hilbert space including
all interacting degrees of freedom. We can therefore consider both the state
of the system $\rho_S (t)$ and of the environment $\rho_E (t)$, as well as the
total state $\rho_{SE} (t)$, assumed to undergo a unitary dynamics. In this
setting, the definition of Markovian reduced dynamics given in
Eq.~(\ref{eq:monotoneq}) can be seen to be connected with a notion of
unidirectional information flow from the system to the environment.

To this aim we can introduce a notion of internal information by identifying
it with the distinguishability between system states according to the
considered quantifier $\mathfrak{S}$, namely
\begin{eqnarray}
  \mathit{I}_{\mathrm{int}} (t) & = & \mathfrak{S} (\rho_S^1 (t), \rho_S^2
  (t)) .  \label{eq:Iint}
\end{eqnarray}
The name internal information stresses the fact that this quantity is
determined by performing measurements on the system only. The existence of a
reduced dynamics is warranted by the choice of a factorized initial condition,
that is
\begin{eqnarray}
  \rho_{SE} (0) & = & \rho_S (0) \otimes \rho_E, 
\end{eqnarray}
with $\rho_E$ a fixed environmental state independent from the system initial
condition. We now observe that Eq.~(\ref{eq:cpcontra}) implies in particular
invariance under a unitary transformation
\begin{eqnarray}
  \mathfrak{S} (U \rho U^{\dag}, U \sigma U^{\dag}) & = & \mathfrak{S} (\rho,
  \sigma),  \label{eq:inv1}
\end{eqnarray}
as well as under the tensor product with respect to a fixed environmental
state
\begin{eqnarray}
  \mathfrak{S} (\rho \otimes \rho_E, \sigma \otimes \rho_E) & = & \mathfrak{S}
  (\rho, \sigma) .  \label{eq:inv2}
\end{eqnarray}
We thus have in particular
\begin{eqnarray}
  \mathfrak{S} (\rho_{SE}^1 (t), \rho_{SE}^2 (t)) & = & \mathfrak{S}
  (\rho_{SE}^1 (0), \rho_{SE}^2 (0)) \nonumber\\
  & = & \mathfrak{S} (\rho_S^1 (0), \rho_S^2 (0)) . 
\end{eqnarray}
We thus introduce the expression
\begin{eqnarray}
  \mathit{I}_{\mathrm{ext}} (t) & = & \mathfrak{S} (\rho_S^1 (0), \rho_S^2
  (0)) - \mathfrak{S} (\rho_S^1 (t), \rho_S^2 (t)),  \label{eq:Iext}
\end{eqnarray}
called external information, complementary to Eq.~(\ref{eq:Iint}) in the sense
that their sum is a constant fixed by the initial distinguishability
\begin{eqnarray}
  \mathit{I}_{\mathrm{int}} (t) + \mathit{I}_{\mathrm{ext}} (t) & = &
  \mathfrak{S} (\rho_S^1 (0), \rho_S^2 (0)) .  \label{eq:Icost}
\end{eqnarray}
The Markov condition Eq.~(\ref{eq:monotoneq}) can now be reformulated as
\begin{eqnarray}
  \mathit{I}_{\mathrm{int}} (t) - \mathit{I}_{\mathrm{int}} (s) & \leqslant &
  0,  \label{eq:monotoneinfoq}
\end{eqnarray}
for any $t \geqslant s$, expressing the fact that the internal information,
namely the capability to distinguish the system states after interaction with
the environment, is steadily decreasing for any pair of initial states. At any
time $t$ the locally available information has diminished with respect to the
initial time, so that to recover the full information at any intermediate time
we would need to perform measurements on all the degrees of freedom, since the
information has been stored outside the system degrees of freedom. If this
happens in a monotonic way, the dynamics is said Markovian according to
Eq.~(\ref{eq:monotoneq}), while a non-Markovian behavior corresponds to a
backflow of information, which becomes again locally retrievable.

\subsection{Information backflow and external information
storage}\label{sect:info-backflow}

We briefly mention another important aspect in this framework for the
description of memory effects, already stressed in
{\cite{Breuer2016a,Campbell2019b}}. Exploiting property Eq.~(\ref{eq:tlike}),
namely the triangle-like inequality, we obtain the following constraint on
possible revivals in the internal information {\cite{Smirne2022b}}
\begin{eqnarray}
  \mathit{I}_{\mathrm{int}} (t) - \mathit{I}_{\mathrm{int}} (s) & \leqslant &
  \phi \circ \phi (\mathfrak{S}(\rho_E^1 (s), \rho_E^2 (s))) \nonumber\\
  &  & + \phi (\mathfrak{S}(\rho_{SE}^1 (s), \rho_S^1 (s) \otimes \rho_E^1
  (s))) \nonumber\\
  &  & + \phi (\mathfrak{S}(\rho_{SE}^2 (s), \rho_S^2 (s) \otimes \rho_E^2
  (s))) . 
\end{eqnarray}
The properties of the function $\phi$, simply corresponding to the identity
function $\phi (x) = x$ for the case in which $\mathfrak{S}$ is a distance
obeying the standard triangle inequality, implies that the three terms at the
r.h.s. are non-negative. A revival in the internal information, corresponding
to a non-Markovian behavior, can therefore only take place if either the same
initial environmental state has differently evolved in correspondence to
different initial system states, or correlations have been established between
system and environment. These conditions, corresponding to the storage of
information outside the system, are necessary in order to have memory effects.
In this respect the property Eq.~(\ref{eq:tlike}) plays a conceptually
relevant role and its verification for entropic distinguishability quantifiers
is crucial for their use in the study of non-Markovian dynamics.

\section{Interaction strength and temperature dependence of trace distance and
entropic distinguishability quantifiers}

To showcase and compare the behavior of different distinguishability
quantifiers in the assessment of the Markovian or non-Markovian behavior of a
quantum dynamics, we consider a physical model of decoherence that covers a
wide variety of physical situations, namely a two-level system interacting
with a collection of bosonic degrees of freedom {\cite{Palma1996a}}. As we
shall show, the advocated notion of memory provides a robust concept,
independently of the specific quantifier considered, with the caveat of
slightly different sensitivity and behaviors with respect to the external
information storage. While the different behavior with respect to composition
properties of the time evolution has been investigated in
{\cite{Settimo2022a}} and the different behavior with respect to the external
information storage in {\cite{Megier2021a,Smirne2022b}}, in the present work
we study the dependence of the non-Markovianity measure on physical parameters
of the model.

\subsection{Decoherence function for the spin-boson model}\label{sec:decohG}

We write the Hamiltonian of the spin-boson model in the form
\begin{eqnarray}
  H & = & H_S + H_E + V,  \label{eq:h}
\end{eqnarray}
with $H_S$ the Hamiltonian of the two-level system with energy gap $\hbar
\omega_0$ and $H_E$ the Hamiltonian of the free bosonic modes. The coupling
term is of the form
\begin{eqnarray}
  V & = & H_S \otimes X,  \label{eq:coupling}
\end{eqnarray}
with $X$ expressed in terms of linear combinations of the bosonic creation and
annihilation operators, so that
\begin{eqnarray}
  {}[H_S, V] & = & 0.  \label{eq:commuta}
\end{eqnarray}
As detailed e.g. in {\cite{Breuer2002}}, the commutator Eq.~(\ref{eq:commuta})
implies that only the coherences of the two-level system are affected, in
particular they are modified according to
\begin{eqnarray}
  \rho_S^{\tmop{offdiag}} (t) & = & \mathe^{- \Gamma (t)}
  \rho_S^{\tmop{offdiag}} (0) .  \label{eq:map}
\end{eqnarray}
The positive function $\Gamma (t)$ starting from zero is called decoherence
function and takes the form
\begin{eqnarray}
  \Gamma (t) & = & \int^{\infty}_0 \mathd \omega J (\omega) \frac{1 - \cos
  (\omega t)}{\omega^2} \coth \left( \frac{\beta}{2} \hbar \omega \right), 
  \label{eq:decohfactorcontinue}
\end{eqnarray}
for the case of a thermal bath of bosonic modes whose coupling to the system
is described by a spectral density $J (\omega)$ {\cite{Weiss1993}}.

\begin{figure}[h]
  \
  
  \resizebox{1\columnwidth}{!}{\includegraphics{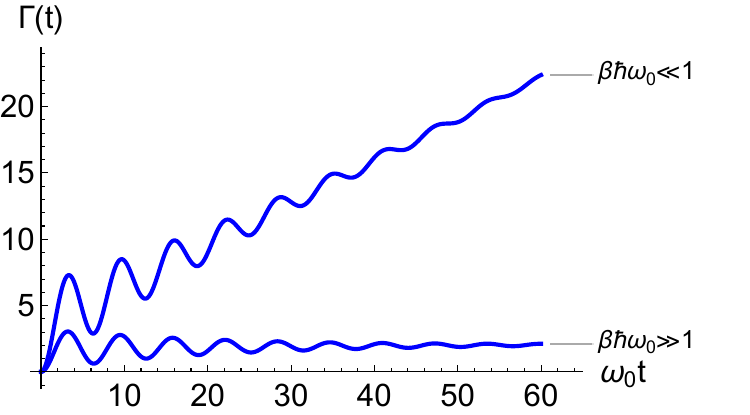}} {\vspace{8pt}}
  \caption{\label{fig:Gamma}Behavior at high and low temperature of the
  decoherence function $\Gamma (t)$ given by Eq.~(\ref{eq:Gammagarg}),
  obtained considering the spectral density of Eq.~(\ref{eq:garg}) and a
  thermal environmental state with inverse temperature $\beta$.}
\end{figure}

In the present treatment following {\cite{Ingold2002a}} we will consider a
spectral density of the form
\begin{eqnarray}
  J (\omega) & = & \kappa \omega_0^2 \frac{\eta \omega}{(\omega^2 -
  \omega_0^2)^2 + \eta^2 \omega^2},  \label{eq:garg}
\end{eqnarray}
where $\omega_0$ is the resonant frequency of the system, $\kappa$ quantifies
the coupling strength, $\eta$ provides the width of the relevant frequency
band {\cite{Garg1985a,Mukamel1995}}. We consider the representation formula
{\cite{Gradshteyn1965}}
\begin{eqnarray}
  \coth \left( \frac{\beta \hbar \omega}{2} \right) & = & \frac{2}{\beta \hbar
  \omega} + \frac{1}{\beta \hbar} \sum_{n = 1}^{+ \infty}
  \frac{\omega}{\omega^2 + \nu_n^2},  \label{eq:matsubara}
\end{eqnarray}
for the hyperbolic cotangent, with $\nu_n = 2 \pi n / \beta \hbar$ the
so-called Matsubara frequencies, so that combining
Eq.~(\ref{eq:decohfactorcontinue}) and Eq.~(\ref{eq:garg}) we come to the
expression
\begin{eqnarray}
  \Gamma (t) & = & \frac{\kappa \pi}{2 \omega^4_0 \Omega} \frac{\sinh (\beta
  \hbar \Omega)}{\cosh (\beta \hbar \Omega) - \cos \left( \frac{\beta \hbar
  \eta}{2} \right) } \nonumber\\
  &  & \times \left\{ \mathe^{- \frac{\eta t}{2}} \left[ \left(
  \frac{\eta^2}{4} - \Omega^2 \right) \cos (\Omega t) - \eta \Omega \sin
  (\Omega t) \right] + \frac{\eta}{2} \omega^2_0 t - \left( \frac{\eta^2}{4} -
  \Omega^2 \right) \right\} \nonumber\\
  &  & + \frac{\kappa \pi}{2 \omega^4_0 \Omega} \frac{\sinh \left(
  \frac{\beta \hbar \eta}{2} \right)}{\cosh (\beta \hbar \Omega) - \cos \left(
  \frac{\beta \hbar \eta}{2} \right) } \nonumber\\
  &  & \times \left\{ \mathe^{- \frac{\eta t}{2}} \left[ \left(
  \frac{\eta^2}{4} - \Omega^2 \right) \sin (\Omega t) + \eta \Omega \cos
  (\Omega t) \right] + \Omega \omega^2_0 t + \eta \Omega \right\} \nonumber\\
  &  & - \eta \frac{\kappa \pi}{\beta \hbar} \sum_{n = 1}^{\infty}
  \frac{1}{\nu_n}  \frac{\mathe^{- \nu_n t} + \nu_n t - 1}{(\nu_n^2 +
  \omega_0^2)^2 - \eta^2 \nu_n^2},  \label{eq:Gammagarg}
\end{eqnarray}
with
\begin{eqnarray}
  \Omega & = & \sqrt{\omega_0^2 - \frac{\eta^2}{4}},  \label{eq:Omega}
\end{eqnarray}
and we consider the so-called underdamped limit in which $\eta / 2 <
\omega_0$. The behavior of this decoherence function $\Gamma (t)$ as a
function of time in the low and high temperature regime is plotted in
Fig.~\ref{fig:Gamma}

We will now study the behavior of the non-Markovianity measure introduced in
Sect.~\ref{sect:markov-measure} for the different distances and entropic
distinguishability quantifiers introduced in Sect.~\ref{sect:quantifiers}.

\subsection{Non-Markovianity measure}

In the considered dephasing model the reduced system dynamics is fixed by the
transformation Eq.~(\ref{eq:map}). This allows to evaluate the expression
$\mathfrak{S} (\rho_S^1 (t), \rho_S^2 (t))$ for relevant choices of
$\mathfrak{S}$ given by $D$, $\sqrt{J}$, $K_{1 / 4}$ and $S_{1 / 4}$. We will
consider an initial pair of system states given by two orthogonal pure states
on the equator of the Bloch sphere. These pairs have the maximal initial
distinguishability according to Eq.~(\ref{eq:normorto}), and are mostly
affected by the dynamics since they both initially have the maximum amount of
coherence. It has further been shown that these pairs maximize the measure
according to Eq.~(\ref{eq:nm-measure}) for the case in which $\mathfrak{S}$ is
the trace distance. We thus have
\begin{eqnarray}
  \rho_S^{1, 2} (0) & \rightarrow & P^{1, 2}_{\varphi} ,
\end{eqnarray}
with $P^{1, 2}_{\varphi}$ projections on the pure orthogonal states
\begin{equation}
  \frac{1}{\sqrt{2}} (|1 \rangle \pm \mathe^{i \varphi} |0 \rangle),
\end{equation}
and we obtain for the trace distance
\begin{eqnarray}
  D (\rho_S^1 (t), \rho_S^2 (t)) & = & \mathe^{- \Gamma (t)}, 
  \label{eq:td-qubit}
\end{eqnarray}
as well as for the distance given by the square root of the Jensen-Shannon
divergence
\begin{eqnarray}
  \sqrt{J (\rho_S^1 (t), \rho_S^2 (t))} & = & \sqrt{1 - H \left( \left\{
  \tfrac{1}{2} (1 + \mathe^{- \Gamma (t)}), \tfrac{1}{2} (1 - \mathe^{- \Gamma
  (t)}) \right\} \right)} .  \label{eq:js-qubit2}
\end{eqnarray}
The other quantifiers can be evaluated starting from the expression
Eq.~(\ref{eq:qre}), exploiting the fact that for the case at hand $[\rho_S^1
(t), \rho_S^2 (t)] = 0$. We finally obtain for the quantum skew divergence
with $\mu = 1 / 4$
\begin{eqnarray}
  S_{1 / 4} (\rho_S^1 (t), \rho_S^2 (t)) & = & \left( \frac{1}{4 \ln 4} +
  \frac{3}{4 \ln (4 / 3)} \right) \left[ 1 - H \left( \left\{ \tfrac{1}{2} (1
  + \mathe^{- \Gamma (t)}), \tfrac{1}{2} (1 - \mathe^{- \Gamma (t)}) \right\}
  \right) \right] \nonumber\\
  &  & - \frac{1}{8 \ln (4)} \left[ \ln \left( 1 - \tfrac{1}{4} \mathe^{- 2
  \Gamma (t)} \right) + \mathe^{- \Gamma (t)} \ln \left( \frac{1 - \frac{1}{2}
  \mathe^{- \Gamma (t)}}{1 + \frac{1}{2} \mathe^{- \Gamma (t)}} \right)
  \right] \nonumber\\
  &  & - \frac{3}{8 \ln (4 / 3)} \left[ \ln \left( 1 - \tfrac{1}{4} \mathe^{-
  2 \Gamma (t)} \right) - \mathe^{- \Gamma (t)} \ln \left( \frac{1 -
  \frac{1}{2} \mathe^{- \Gamma (t)}}{1 + \frac{1}{2} \mathe^{- \Gamma (t)}}
  \right) \right], \nonumber\\
  &  & 
\end{eqnarray}
and for the Holevo skew divergence with $\mu = 1 / 4$
\begin{eqnarray}
  K_{1 / 4} (\rho (t), \sigma (t)) & = & \frac{8}{4 + 6 \ln (4 / 3)} \left[ 1
  - H \left( \left\{ \tfrac{1}{2} (1 + \mathe^{- \Gamma (t)}), \tfrac{1}{2} (1
  - \mathe^{- \Gamma (t)}) \right\} \right) \right] \nonumber\\
  &  & - \frac{4}{4 + 6 \ln (4 / 3)} \ln \left( 1 - \frac{1}{4} \mathe^{- 2
  \Gamma (t)} \right) \nonumber\\
  &  & + \frac{2 \mathe^{- \Gamma (t)}}{4 + 6 \ln (4 / 3)} \ln \left( \frac{1
  - \frac{1}{2} \mathe^{- \Gamma (t)}}{1 + \frac{1}{2} \mathe^{- \Gamma (t)}}
  \right) . 
\end{eqnarray}
\subsubsection{Temperature and coupling strength dependence}\label{sect:TK}

We can now exploit the explicit time-dependent expressions for the decoherence
function and for the different distinguishability quantifiers to investigate
the temperature and coupling dependence of the non-Markovianity measure.
Making reference to Eq.~(\ref{eq:nm-sum}) and Eq.~(\ref{eq:td-qubit}) we have
\begin{eqnarray}
  \Mu (\Phi, D, P^{1, 2}_{\varphi}) & = & \sum_n  [\mathe^{- \Gamma
  (t^n_f)} - \mathe^{- \Gamma (t^n_i)}],  \label{eq:measure-td}
\end{eqnarray}
while considering Eq.~(\ref{eq:js-qubit2}) we obtain
\begin{eqnarray}
  \Mu (\Phi, \sqrt{J}, P^{1, 2}_{\varphi}) & = & \sum_n  \left[
  \sqrt{1 - H \left( \left\{ \tfrac{1}{2} (1 + \mathe^{- \Gamma (t^n_f)}),
  \tfrac{1}{2} (1 - \mathe^{- \Gamma (t^n_f)}) \right\} \right)} \right.
  \nonumber\\
  &  & \left. \hspace{3em} - \sqrt{1 - H \left( \left\{ \tfrac{1}{2} (1 +
  \mathe^{- \Gamma (t^n_i)}), \tfrac{1}{2} (1 - \mathe^{- \Gamma (t^n_i)})
  \right\} \right)} \right],  \label{eq:measure-js}
\end{eqnarray}
where the pairs $\{ t^n_i, t^n_f \}_n$ denote the time windows in which the
distinguishability quantifiers shows revivals. Similar but more cumbersome
expressions can be written for $K_{1 / 4}$ and $S_{1 / 4}$. All of these
expressions provide monotonically increasing functions of $\Gamma$, so that
the pairs $\{ t^n_i, t^n_f \}_n$ are determined by the time regions in which
$\Gamma (t)$ decreases, that can be determined numerically.

\begin{figure}[h]
  \resizebox{1\columnwidth}{!}{\includegraphics{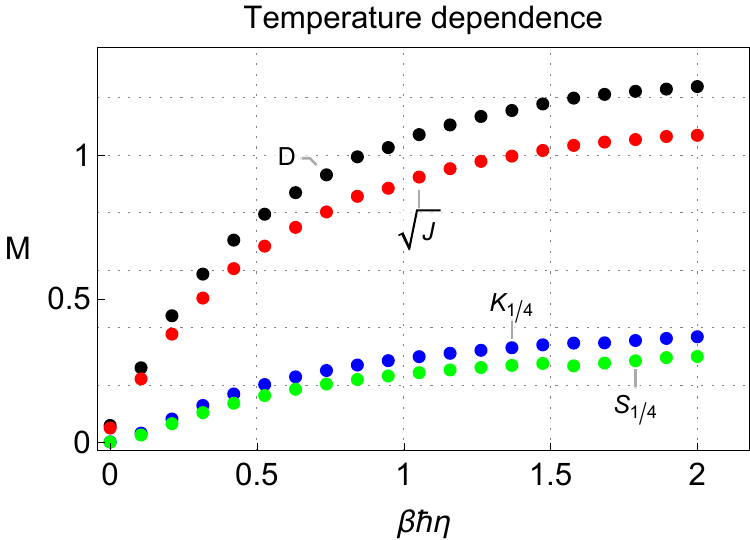}}
  {\vspace{8pt}}
  \caption{\label{fig:beta}Plot of the non-Markovianity measure for the
  spin-boson dephasing model in its dependence on the inverse temperature, for
  the different quantifiers $D$, $\sqrt{J}$, $K_{1 / 4}$ and $S_{1 / 4}$.}
\end{figure}

\begin{figure}[h]
  \resizebox{1\columnwidth}{!}{\includegraphics{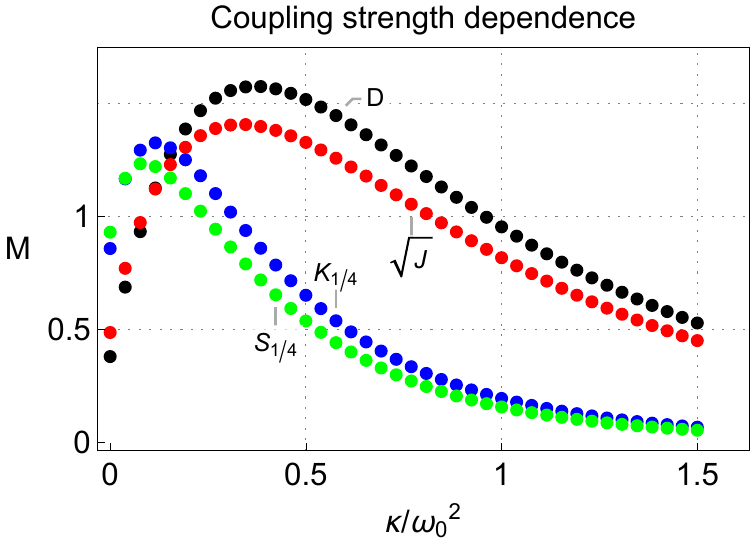}}
  {\vspace{8pt}}
  \caption{\label{fig:kappa}The same as in Fig.~\ref{fig:beta}, but
  considering the coupling strength dependence of the non-Markovianity
  measure.}
\end{figure}

The behavior of these measures as a function of inverse temperature $\beta$
and coupling strength $\kappa$ are plotted in Fig.~\ref{fig:beta} and
Fig.~\ref{fig:kappa} respectively. For the case of the temperature dependence
they all show an increase of non-Markovianity for decreasing temperature. The
two distances $D$, $\sqrt{J}$ have a very similar behavior and higher
sensitivity with respect to the two divergences $K_{1 / 4}$ and $S_{1 / 4}$.
The Holevo skew divergence $K_{1 / 4}$ further shows a slightly stronger
temperature dependence when compared with the quantum skew divergence $S_{1 /
4}$. Investigating the coupling dependence strength all quantifiers point to a
non-monotonic behavior of the measure, initially increasing and later
vanishing for strong enough coupling. Also in this case the two distances $D$,
$\sqrt{J}$ behave similarly and exhibit a higher sensitivity with respect to
the two divergences $K_{1 / 4}$ and $S_{1 / 4}$. Again the Holevo skew
divergence $K_{1 / 4}$ features a slightly more marked coupling strength
dependence with respect to the quantum skew divergence $S_{1 / 4}$.

\section{Conclusions}

We have considered a general framework for the characterization of quantum
non-Markovian dynamics, that is based on the behavior in time of the
distinguishability between system states obtained starting from different
initial conditions. The framework includes both distances and divergences.
After a brief discussion of the physical motivation behind the approach, we
have considered the behavior of distances and divergences in the study of a
general physical model of decoherence. We investigate in particular the amount
of non-Markovianity associated to the model in the dependence on physical
parameters such as temperature and coupling strength. The behavior of two
distances, namely trace distance and square root of the Jensen-Shannon
divergence, is compared with the one of two divergences, namely Holevo and
quantum skew divergence. It appears that the approach is robust with respect
to the choice of quantifier, in that they all exhibit the same dependence on
the physical parameters of the model. At the same time distances are more
sensitive than divergences in their dependence on these parameters.

\section*{Acknowledgments}

The work was partially supported by the Italian MIUR under PRIN 2022.

\end{document}